\documentclass[twocolumn,prl,preprintnumbers,amsmath,amssymb]{revtex4}

\usepackage{graphicx}
\usepackage{dcolumn}
\usepackage{bm}
\usepackage{color}

\begin{document}


\title{Tachyon Condensation Due to Domain-Wall Annihilation in Bose-Einstein Condensates
}

\author{Hiromitsu Takeuchi$^1$}
\email[]{hiromitu@hiroshima-u.ac.jp}
\author{Kenichi Kasamatsu$^2$}
\author{Makoto Tsubota$^3$}
\author{Muneto Nitta$^4$}

\affiliation{%
$^1$ Graduate School of Integrated Arts and Sciences, Hiroshima University, Kagamiyama 1-7-1, Higashi-Hiroshima 739-8521, Japan\\
$^2$ Department of Physics, Kinki University, Higashi-Osaka 577-8502, Japan\\
$^3$ Department of Physics and The Osaka City University Advanced Research Institute for Natural Science and Technology (OCARINA), Osaka City University, Sumiyoshi-ku, Osaka 558-8585, Japan \\
$^4$ Department of Physics and Research and Education Center for Natural Sciences, Keio University, Hiyoshi 4-1-1, Yokohama, Kanagawa 223-8521, Japan
}%


\date{\today}

\begin{abstract}
We show theoretically that a domain-wall annihilation in two-component Bose-Einstein condensates causes tachyon condensation accompanied by spontaneous symmetry breaking in a two-dimensional subspace.
Three-dimensional vortex formation from domain-wall annihilations is considered a kink formation in subspace.
Numerical experiments reveal that the subspatial dynamics obey the dynamic scaling law of phase ordering kinetics.
This model is experimentally feasible and provides insights into how the extra dimensions influence subspatial phase transition in higher-dimensional space.
 \end{abstract}


\maketitle

\begin{figure*}[thtb] \centering
  \includegraphics[width=1. \linewidth]{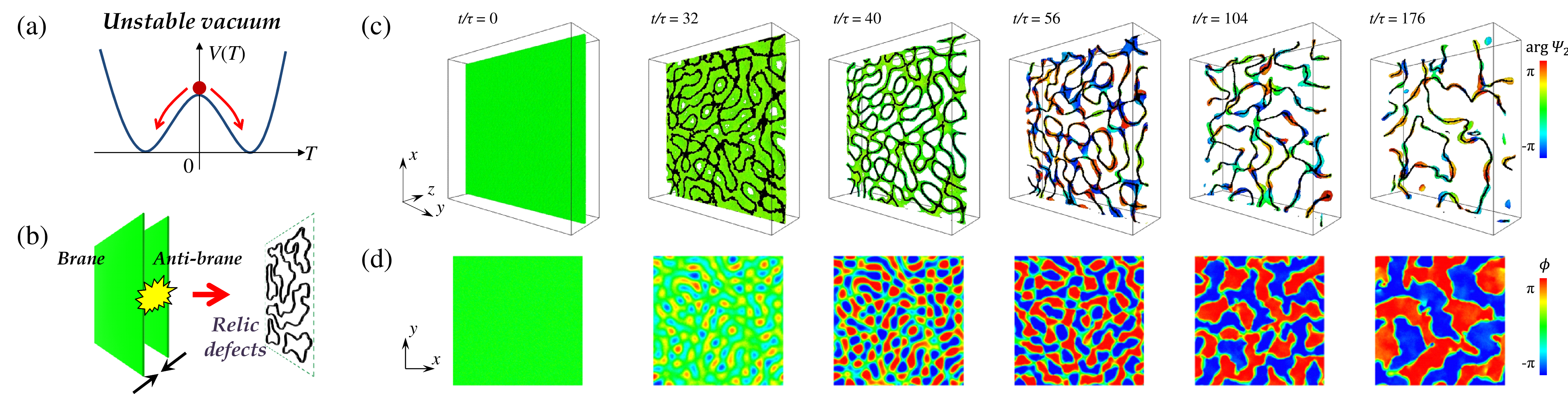}
  \caption{
 (a) Schematic diagrams of tachyon condensation.
 (b) Schematic diagrams of an annihilation of a brane pair and the relic lower-dimensional objects.
(c) Dynamics of a domain wall pair annihilation in a numerical experiment for $\nu=0.84$ and $\Delta \Theta=\pi$.
 Panels show the time development of the isosurface $|\Psi_1|=|\Psi_2|$.
 Black curves represent the cores of vortices in the $\Psi_1$-component (not shown for $t/\tau=0$). The phase $\arg\Psi_1$ strongly fluctuates in the region $|\Psi_1|\sim 0$ between the branes, where there are many vortices with little vorticity of mass current
 ($t/\tau=32$). The initial fluctuation grows into a meshed structure ($t/\tau=40$).
The structure yields serpentine vortices, which trap the $\Psi_2$-component along the vortex cores. The $\Psi_2$-component propagates along the cores and causes varicose oscillations of the isosurface. Reconnections between vortices accidentally create vortex rings or disk-shaped density pulses, which escape to the bulk.
(d) Panels show the time development of the projected field $\phi(x,y,t)$ of Eq.~\eqref{eq:phi}, calculated from the results of (c).
The box sizes are $80\xi\times 80\xi\times 25.6\xi$ for (c) and $102.4 \xi \times 102.4 \xi$ for (d).
 }
\label{fig:annihilation}
\end{figure*}

A tachyon is a hypothetical superluminal particle that violates causality in special relativity. Most physicists deny its existence because it is inconsistent with the known laws of physics.
However, in quantum field theories, a tachyon {\it field} can exist due to the instability of quantum vacuum. Here, the `instability' means that the state is at a maximum of an effective potential $V(T)$ for the tachyon field $T$ [Fig.~\ref{fig:annihilation}(a)]. The tachyon field grows exponentially with time and rolls down toward a minimum of the potential as the true vacuum. This process is called {\it tachyon condensation} \cite{Sen:2005}.


Tachyon condensation is a key concept used for describing the dynamics of
string theory,
a promising candidate for a `theory of everything' that describes all fundamental forces and forms of matter in nature
\cite{Polchinski:1998}.
In this theory, a tachyon exists in a system containing
a Dirichlet(D)-brane and an anti-D-brane, where the former is an extended solitonic object \cite{Polchinski:1998}
and the latter its anti-object. 
The two annihilate in a collision similarly to particles--antiparticles annihilation in a collision. The annihilation process is described by tachyon condensation,
and the system falls into the true vacuum after complete annihilation \cite{Sen:2005}.

A remarkable application of tachyon condensation
is in brane cosmology \cite{Dvali:2001,Langlois:2002, Quevedo:2002, McAllister:2008}, in which the Big Bang is hypothesized to occur as a result of a collision of a brane and an anti-brane.
After this collision, lower-dimensional branes remain as relics of tachyon condensation \cite{Sen:2005}, which correspond to cosmic strings in brane cosmology \cite{Jones:2002,Sarangi:2002,Dvali:2004} [see Fig.~\ref{fig:annihilation}(b)].
This situation resembles conventional phase transitions accompanied by spontaneous symmetry breaking (SSB), resulting in the formation of topological defects via the Kibble-Zurek mechanism \cite{Kibble:1976,Zurek:1985}.
This mechanism produces topological defects in the early Universe due to phase transitions \cite{Vilenkin:1994}, which has been tested in several condensed matter systems
\cite{Hendry:1994,Bowick:1994,Bauerle:1996,Ruutu:1996,Carmi:2000,Maniv:2003,Monaco:2006,Sadler:2006,Weiler:2008}.
In contrast, tachyon condensation as a SSB phenomenon has not yet been well understood. Because it may lead to defect nucleation in a restricted lower-dimensional subspace, the dynamics should be affected by the degree of freedom associated with the extra dimension. However, the influence of the extra dimension has never been discussed, partly because such phenomena are absent in actual systems.

\begin{figure} [thbp] \centering
  \includegraphics[width=1. \linewidth]{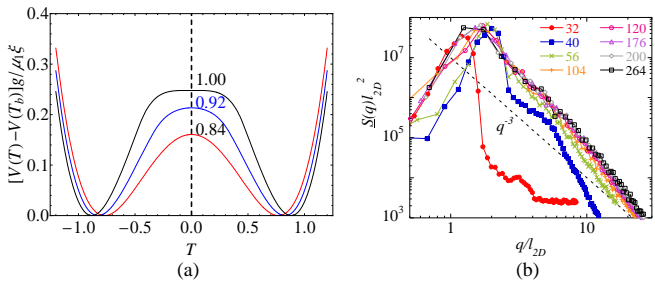}
  \caption{
(a) Plots of the tachyon potential with $\Delta\Theta=\pi$ for $\nu=0.84$, $0.92$, and $1.00$.
 The potential $V(T)$ takes its minimum value $V(T_b)$ at $T=\pm T_b$.
(b) Structure factor of $\phi$ in the numerical experiment [Fig.~\ref{fig:annihilation}(d)].
 The factor $\underline{S}(q,t)$ is the average of $S({\bm q},t)$ over directions of ${\bm q}$.
 The dashed line is the expected law, $S(q)\propto q^{-3}$ for $q/l_{\rm 2D}\gg 1$, which is not observed in the small system.
}
\label{fig:Tpotential}
\end{figure}

Here, we provide a groundbreaking system to
tackle this problem, using atomic Bose--Einstein condensates (BECs).
Tachyon condensation is simulated by 
considering vortex formations from a pair-annihilation of domain walls, i.e. branes, in binary BECs.
 This system is advantageous in that
 we can theoretically and experimentally address the nonlinear dynamics of branes, such as defect nucleation and subsequent dynamics,
 which is difficult in string theory.
 Anderson {\it et al.} \cite{Anderson:2001} observed the creation of vortex rings via the dynamic (snake) instability of a dark soliton in two-component BECs, where the nodal plane of a dark soliton in one component was filled with the other component, and then, the filling component was selectively removed using a resonant laser beam. Recently, we proposed that domain walls in phase-separated two-component BECs correspond to D-branes in the sense that vortex lines (strings) can terminate on them \cite{Kasamatsu:2010}.
Therefore, the experiment in Ref. \cite{Anderson:2001} may be interpreted as the demonstration of defect formations via brane annihilation \cite{Takeuchi:2011},
although those phenomena have never been understood as SSB phenomena in a restricted lower-dimensional subspace.
 In this Letter, we theoretically show that a domain wall annihilation causes spontaneous $Z_2$ symmetry breaking and phase ordering dynamics in the two-dimensional subspace,
where a tachyon field is introduced by {\it projecting} the original order parameters {\it onto} the branes in three-dimensional space. 
 Our theory is justified by demonstrating the scaling law of phase ordering kinetics \cite{Bray:1994} in numerical experiments  [see Fig.~\ref{fig:annihilation}(d) and Fig.~\ref{fig:Tpotential}(b)].
Although the analogue of the brane annihilation was simulated experimentally on the AB phase boundary of superfluid $^3$He \cite{Bradley:2008},
 its theoretical explanation remains lacking.
 This work provides the first theory of brane-annihilation phenomena in condensed matter systems.

We consider two-component BECs, which consist of condensations of two distinguishable Bose particles.
Two-component BECs are well described by two complex order parameters, $\Psi_j~(j=1, 2)$, in the Gross-Pitaevskii model at zero temperature \cite{Pethick:2002}. The order parameters in an uniform system obey the action
${\cal S}=\int dt \int d^3x\left( i\hbar \sum_j \Psi_j^*\partial_t \Psi_j-{\cal K}-{\cal V} \right)$
with kinetic energy density
${\cal K}=\sum_j\frac{\hbar^2}{2m_j}|{\bf \nabla}\Psi_j|^2$
and potential energy density
${\cal V}=\sum_j \left( \sum_k \frac{g_{jk}}{2}|\Psi_k|^2-\mu_j \right)|\Psi_j|^2$.
Here, we have expressed the coupling constant $g_{jk}=2\pi\hbar^2 a_{jk}/m_{jk}~(j,k=1,2)$ with the reduced mass $m_{jk}^{-1}=m_j^{-1}+m_k^{-1}$ and the $s$-wave scattering length $a_{jk}$ between atoms in the $\Psi_j$- and $\Psi_k$-components. The chemical potential $\mu_j$ is introduced as the Lagrange multiplier for the conservation of the norm $N_j=\int  d^3x |\Psi_j|^2={\rm const}.$, which gives the particle number of the $\Psi_j$-component. We consider strongly segregated BECs and set the parameters as $m_1=m_2\equiv m$, $g_{11}=g_{22}\equiv g$, $ \mu_2/\mu_1\equiv\nu$, and $ g_{12}/g\equiv\gamma=2$;
this parameter setting is experimentally feasible, {\it e.g.} Ref.~\cite{Papp:2008}.
The time and length scales of our system are characterized by $\tau\equiv \hbar/\mu_1$ and $\xi \equiv \hbar/\sqrt{m \mu_1}$, respectively.

Let us consider the annihilation of a domain wall (brane) at $z=-R/2$ and an anti-domain wall (anti-brane) at $z=R/2$ perpendicular to the $z$-axis, between which the $\Psi_2$-component is sandwiched by the two domains occupied with the $\Psi_1$-component. We define the inter-brane distance $R$ as the distance between the two planes defined by $|\Psi_1|=|\Psi_2|$. The distance $R$ increases with $N_2$, which is controllable experimentally \cite{Anderson:2001}. Because the `penetration' of the amplitude $|\Psi_1|$ ($|\Psi_2|$) decays exponentially with distance into the $\Psi_2$-($\Psi_1$-)domain, the short-range interaction between the branes works effectively only when $R$ is comparable to the `penetration depth', or the brane thickness, and then, the annihilation process can start substantially. The trivial process of pair annihilation is that the branes collide to leave the trivial state $\Psi_1={\rm const.}$.
However, the annihilation processes become nontrivial depending on the phase difference $\Delta \Theta \equiv \arg\Psi_1(z \gg R/2)-\arg\Psi_1(z \ll -R/2)$.

Since the essential mechanism of the nontrivial annihilation has been discussed partly in Ref.~\cite{Takeuchi:2011},
we explain it briefly here \cite{Supp}.
 To capture the essence of the annihilation process,
we assume that two branes with large $R (\gg \xi)$ are brought
rapidly to a small distance $R\lesssim \xi$ but the two $\Psi_1$-domains are disconnected at $t=0$ \cite{Foot3}.
 In the annihilation process, junctions connecting the two $\Psi_1$-domains emerge in various places on the $x$-$y$ plane around $z=0$.
For $\Delta\Theta\neq 0$, a junction causes a superfluid current along the $z$-axis with a current velocity
 $v_\bot \sim \frac{\hbar}{m}\Delta\Theta/R>0$ or $v_\bot  \sim\frac{\hbar}{m}(\Delta\Theta-2\pi)/R<0$.
If the current velocities through the two neighbouring junctions are parallel, the annihilation is completed between the junctions.
 On the other hand, the two junctions with opposite velocities leave a single-quantum vortex in the $\Psi_1$-component, 
where the $\Psi_2$-component is trapped in the vortex cores so that $N_1$ and $N_2$ are conserved.

Although the growth rates of junctions with $v_\bot >0$ and $v_\bot <0$ are generally different, they are statistically equivalent for $\Delta\Theta = \pi$.
Then, the junctions grow in a random fashion from initial random fluctuations, and vortices emerge as serpentine curves along the boundary between the two opposite junctions.
These scenarios are demonstrated numerically as shown in Fig.~\ref{fig:annihilation}(c) \cite{Supp}. 
 The snake instability observed by Anderson {\it et al.} corresponds to the coincident limit  of the two branes ($R\to 0$) with $\Delta\Theta=\pi$, where the vortex ring nucleation results from the spherical geometry of the external potential \cite{Anderson:2001}. More generally, we can regard the inter-brane distance $R$ and the phase difference $\Delta \Theta$ as two parameters to characterize the dynamics of the brane annihilation.

To describe the annihilation process systematically, we construct an effective field theory parametrized with $R$ and $\Delta\Theta$. Tachyon condensation in string theory is explained by introducing a growing field, that is, the tachyon field, in the lower-dimensional space spanned by the coordinates along the branes \cite{Sen:2005}.
In a similar manner, we consider an effective tachyon field $T(x,y,t)$ as a real scalar field in a two-dimensional space parametrized by $x$ and $y$. 
On the basis of the vortex formation mechanism explained above, we introduce the variational ansatz,
\begin{eqnarray}
&&
\hspace{-.8 cm}
\Psi_1 =\sqrt{\mu_1/g}\left[\sqrt{\tanh^2(z/\xi_\bot)e^{\pm i \Delta \Theta}}+T{\rm sech}^2(z/\xi_\bot)\right],
\label{eq:ansatz1}
\\
&&
\hspace{-.8 cm}
\Psi_2 =\sqrt{n_2(T)} {\rm sech}(z/\xi_\bot),
\label{eq:ansatz2}
\end{eqnarray}
where $T(t=0)=0$ and the sign $\pm$ in Eq. \eqref{eq:ansatz1} takes $+$ ($-$) sign for $z<0$ ($z>0$) \cite{Foot4}.
 This ansatz describes the annihilation starting from a small inter-brane distance, where the population of the $\Psi_2$-component is small between the branes at $t=0$. The parameters $n_2^0\equiv n_2(t=0) (\geq 0)$ and $\xi_\bot$ are determined to minimize the energy $\int d^3x ({\cal K}+{\cal V})$ at $t=0$. The ansatz at $t=0$ reasonably reduces to the solution of a dark soliton with $\xi_\bot \to \xi$ and $R \to 0$ for $n_2^0\to 0$ and $\Delta \Theta=\pi$. The growth of the tachyon field from $T=0$ into $T>0$ ($T<0$) induces a superflow with $v_\bot >0$ ($v_\bot<0$).
Although the dynamics may be described more precisely with additional variational parameters,
our simplest ansatz is enough to capture the essence of the vortex formation dynamics.

The effective potential for the tachyon field $T$, namely the tachyon potential $V$, is defined as a function of $T$, $V(T)=\int^{+\infty}_{-\infty}dz({\cal K}+{\cal V})$ with $\partial_x\Psi_j=\partial_y\Psi_j=0$. We assume that the density parameter $n_2(T) \geq 0$ is determined so as to minimize $V$. It is straightforward to obtain the form
\begin{eqnarray}
V=\frac{\mu_1^2\xi_\bot}{g}\sum_{n=0}^4 F_n{T}^n,
\label{TachyonPotential}
\end{eqnarray}
where $F_n=A_n(\Delta\Theta,\nu,\gamma)+B_n(\Delta\Theta,\nu,\gamma)\theta(n_2)$ with a step function, $\theta(n_2>0)=1$ and $\theta(n_2\leq 0)=0$.
The contribution from the gradient $\bm{\nabla}_\parallel \equiv (\partial_x,\partial_y)$ along the coordinates parallel to the brane is calculated similarly. The original energy is then reduced to the form
 \begin{eqnarray}
E_{\rm 2D}=\int dx dy \left[
G(T ) (\xi \bm{\nabla}_{\parallel}T )^2
+V(T)
\right].
\label{eq:E2D}
\end{eqnarray}
Here, the coefficient $G(T)>0$ in the gradient term depends on $T$ (see Ref.~\cite{Supp} for details).

Equation \eqref{eq:E2D} represents the effective energy for the field $T$ in the `projected-2D' system. The potential $V$ is symmetric $V(-T)=V(T)$ for $\Delta \Theta = \pi$,
because the coefficients $F_1$ and $F_3$ are proportional to $\cos (\Delta \Theta / 2)$.
The $Z_2$ symmetry comes from the degeneracy of the kinetic energy $\propto v_\bot^2 \sim (\hbar\pi/mR)^2$ for $\pm T$ owing to a symmetry in the initial configuration; ${\rm Re}\Psi_1(t=0) =-{\rm Re}\Psi_1(t=0)$ with $\Delta \Theta = \pi$.
In terms of relativistic quantum field theory,
 we find a particle-like state with a mass $m_T\equiv \sqrt{V''(T=0)/2}=\mu_1\sqrt{\xi_\bot F_2/g}$ by expanding the potential $V$ around $T=0$.
For $F_2>0$, this describes a conventional particle; however, for $F_2<0$, this describes a particle with purely imaginary mass $m_T^2<0$, {\it i.e.} a tachyon. The existence of a tachyon implies instability of the system, because a tachyon rolls down from the potential maximum at $T=0$ toward a potential minimum with $T>0$ or $T<0$ \cite{Foot1}. Figure~\ref{fig:Tpotential}(a) shows the tachyon potential $V(T)$ for $\Delta\Theta=\pi$ and $\gamma=2$. The coefficient $F_2$ increases with $\nu=\mu_2/\mu_1$, and the tachyon potential is convex upward around $T=0$ for $\nu \leq 1$. Because $R$ is an increasing function of $\nu$, the instability becomes stronger as the inter-brane distance $R$ decreases.

The tachyon field $T$ is an analogue of the order-parameter field, {\it e.g.} the magnetization density, of a ferromagnetic system \cite{Landau:1980StatisticalPhysics} in a continuum description.
The rolling tachyon corresponds to the spontaneous $Z_2$-symmetry breaking from the zero `magnetization' $T=0$ toward a macroscopic `magnetization' $T>0$ or $T<0$ in the ordered phase for $\Delta\Theta =\pi$ \cite{Foot2}.
 In this sense, the inter-brane distance $R$ and the phase difference $\Delta\Theta$ ($\neq \pi$) play the roles of the `temperature' and `external magnetic field', respectively. The coefficient $F_2$ increases with the `temperature' $R$, implying that the instability becomes weak for a small inter-brane interaction for large $R$. Because the inter-brane interaction decays exponentially with $R$ for large distance and the instability vanishes precisely for $R\to \infty$, the infinity distance may correspond to the `transition temperature'. Although  $F_2$ reached zero for finite $R$ in our effective model,
the tachyon potential $V(T)$ well reflects the nature of the instability for small $R$.
 On the other hand, the field $T$ feels the `magnetic field' for $0\leq \Delta \Theta < \pi$ ($\pi < \Delta \Theta \leq 2\pi$), and then, `magnetization' $T>0$ ($T<0$) is energetically favourable.
 In the original 3D system, this is energetically explained from the difference in the kinetic energy induced by the rolling tachyon, $\propto v_\bot^2 \sim \left(\frac{\hbar\Delta\Theta}{mR}\right)^2$ for $T>0$ and $\propto v_\bot^2 \sim \left[\frac{\hbar(\Delta\Theta-2\pi)}{mR}\right]^2$ for $T<0$.

 A vortex between junctions with $v_\bot >0$ and $v_\bot <0$ is considered a kink between the regions with $T>0$ and $T<0$ in the projected-2D space.
 The validity of the effective theory is confirmed by evaluating the defect nucleation rate.
The effective theory neglects the transfer of particles in the $x$- and $y$-directions in the defect formation process, violating the law of particle number conservation.
 However, the violation influences little on the rate evaluation for small $R$.
According to the evaluations with several analytical and numerical methods in Ref.~\cite{Supp},
 the rate consistently decreases with $R$.

Finally, we provide the most important evidence showing
 that the field $T$ behaves actually as a two-dimensional order parameter. According to the scaling law in the phase ordering kinetics \cite{Bray:1994}, the spatial structure of the order parameter is characterized by a single length scale, {\it i.e.} the mean inter-defect distance, after a rapid quench from the disordered phase into the ordered phase. In the simulations of the 3D system, the density of kinks in the field $T$ is calculated from the line density $l_{\rm 2D}$ of the projection of vortices onto the $x$-$y$ plane by assuming that there is no overlap between the projection lines of different vortices. To visualize the dynamics of $T$, we introduce the projected field
\begin{eqnarray}
\phi(x,y,t) \equiv \frac{m}{\hbar}\int^{+\infty}_{-\infty} dz v_\bot({\bf r},t)
\label{eq:phi}
\end{eqnarray}
with the superfluid current velocity $v_\bot=\frac{\hbar\sum_j|\Psi_j|^2 \partial_z \arg\Psi_j}{m\sum_k|\Psi_k|^2}$. Far from kinks in the projected-2D space, we have $\Psi_2 \sim 0$ and then both fields $\phi$ and $T$ are constant, $\phi(x,y) \sim \pm \pi$ and $T(x,y)=\pm T_b$, where $V(T)$ takes its minimum value at $T=\pm T_b$.
 The nonzero $\Psi_2$-component around the kink cores could yield different spatial dependences of the two fields. Thus, a spatial structure of the projected field $\phi(x,y,t)$ represents that of $T(x,y,t)$ on a length scale larger than the kink width $\zeta$, which is sufficient for the following analysis.

If the field $\phi$ obey the scaling law, its structure factor $S({\bm q},t)$ for the wavenumber $|{\bm q}|\ll 1/\zeta$ is written with a time-independent function ${\cal F}$ as ${\cal F}(q/l_{\rm 2D})=S({\bm q},t)l_{\rm 2D}^2$.
Here, the scaling form is expected to follow the universal law $S({\bm q},t)\sim (l_{\rm 2D}/q)^{d+1}/l_{\rm 2D}^d$, known as the Porod law, with the spatial dimension $d=2$ for $q/l_{\rm 2D}\gg 1$ \cite{Bray:1994}.


Figure~\ref{fig:Tpotential}(b) shows scaling plots of $S({\bm q},t)$ for the numerical experiment of Fig. \ref{fig:annihilation}(c). The dynamics of $\phi$ in Fig. \ref{fig:annihilation}(d) resembles ferromagnetic relaxations after rapid quenching.
 In fact, the scaling plots almost coincide with each other after the domain structures of $\phi$ become clear.
 The scaling behaviour is seen from the similarity between the patterns of $t/\tau=104$ and $t/\tau=176$ in Fig.~\ref{fig:annihilation}(d). These facts show the domain-wall annihilation is regarded as phase ordering in the projected-2D space.

The time dependence of $l_{\rm 2D}$ contains a statistical information.
The density $l_{\rm 2D}$ follows a power law, $1/l_{\rm 2D}\propto t^{1/2}$ \cite{Supp},
 which indicates that the projected-2D system is dissipative.
 The dissipation may come from some degrees of freedom, neglected in the effective theory, such as the motions of the $\Psi_2$-component along the vortex core, emissions of vortex rings, and density pulses to the extra dimension induced by vortex reconnection [see Fig.~\ref{fig:annihilation} (c) and the caption].
It is interesting that strings stretched between the brane and the anti-brane cause a complex-scalar tachyon field in string theory \cite{Sen:2005}
while the real-scalar field comes from fluctuating fields in the wall-anti-wall background in our system.
 If vortices are stretched between the domain walls,
a `vorton', a point-like defect in 3D, can be nucleated due to superflow of the $\Psi_2$-component along the vortex core \cite{Nitta:2012}.
These issues will be discussed elsewhere.


Our proposal gives the first realistic example of
non-relativistic tachyon condensation due to the brane annihilation phenomena.
Because the instability is essentially caused by the spatially inhomogeneous junctions between two domains described with the same order parameter, the phenomenon may be generalized to a problem of a sudden connection between two media in the same ordered phase. Therefore, it is expected that our theory can be extended to various condensed matter systems. The $^3$He-brane experiment \cite{Bradley:2008} should be retested from the viewpoint proposed in this work. Because techniques for detecting the vortex line density in the $^3$He-system are well developed, some statistical information about the brane annihilation could be observed from the power-law decay of the defect density.


We are grateful to M. Kobayashi for useful discussions, and to N. Hatakenaka for comments on the manuscript.
The authors thank the Supercomputer Center, Institute for Solid State Physics,
University of Tokyo for the facilities and the use of the SGI Altix ICE 8400EX
(and/or NEC SX-9).
This work was supported by KAKENHI from JSPS 
(Grant Nos. 21340104, 21740267 and 23740198).
This work was also supported
by the ``Topological Quantum Phenomena'' 
(Nos. 22103003 and 23103515)
Grant-in Aid for Scientific Research on Innovative Areas 
from the Ministry of Education, Culture, Sports, Science and Technology 
(MEXT) of Japan.


\section*{SUPPLEMENTAL MATERIAL}

\subsection*{Vortex formations from brane annihilation}

\begin{figure*} [hbpt] \centering
  \includegraphics[width=1. \linewidth]{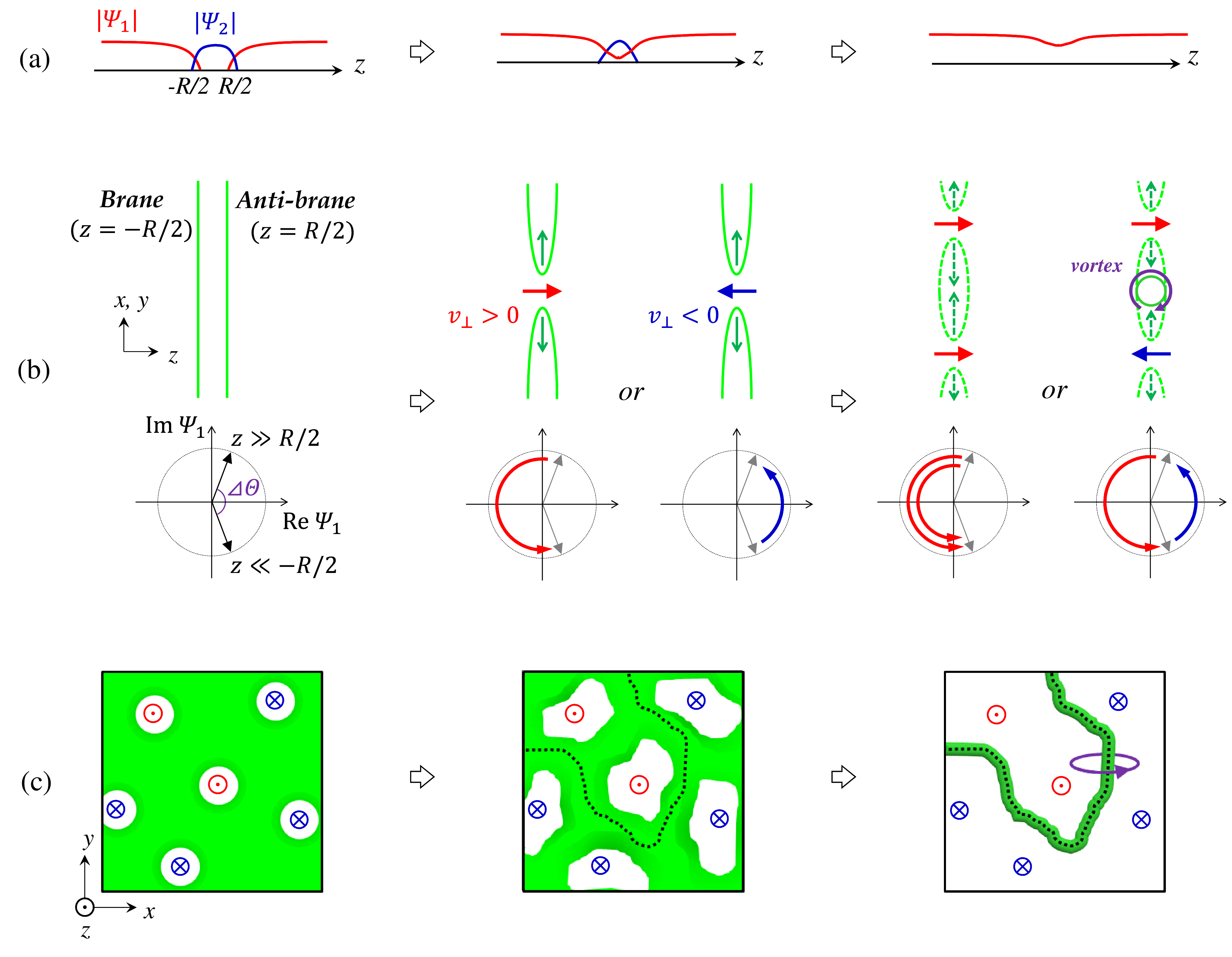}
  \caption{
Schematic diagrams of vortex formation in brane annihilation process.
(a), One-dimensional diagram of the order parameter amplitudes $|\Psi_1|$ and $|\Psi_2|$. The two $\Psi_1$-domains are separated by the $\Psi_2$-domain (left), and then, a junction is formed between the two domains (centre), leading to annihilation (right).
(b), Two-dimensional diagram of the two colliding branes (planes defined by $|\Psi_1|=|\Psi_2|$).
 The lower figures represent the phase $\arg \Psi_1(z)$ and its change of direction between $z \ll -R/2$ and $ z \gg R/2$ in the configuration space.
 The phase difference $\Delta\Theta$ between the two domains (left) induces a superfluid current through the junction with direction $v_\bot>0$ ($\rightarrow$) or $v_\bot<0$ ($\leftarrow$) (centre). Between the two neighbouring junctions, the annihilation finishes or a single-quantum vortex is left (right).
(c), Three-dimensional diagram viewed from the positive direction of the $z$-axis.
 For $\Delta\Theta = \pi$,
 an initial fluctuation develops into a random distribution of junctions with $v_\bot>0$ ($\odot$) or $v_\bot<0$ ($\otimes$) (left), causing a meshed structure (centre). A serpentine vortex line is formed along a boundary (dotted curve) between two junctions with opposite velocity (right).
}
\label{fig:Schematic}
\end{figure*}
Figure~\ref{fig:Schematic} summarizes the brane--anti-brane annihilation and resulting vortex formation in our system.
The trivial process of pair annihilation is that the branes collide to leave the trivial state $\Psi_1={\rm const.}$ by pushing away the $\Psi_2$-component, as shown in Fig.~\ref{fig:Schematic}(a).
However, the annihilation processes become nontrivial depending on the phase difference $\Delta \Theta \equiv \arg\Psi_1(z \gg R/2)-\arg\Psi_1(z \ll -R/2)$.

 To capture the essence of the nontrivial domain wall annihilation, we consider that the brane and the anti-brane are brought close to each other at a small distance $R$ at $t=0$. In the annihilation process, junctions connecting the two $\Psi_1$-domains emerge in various places between the branes by pushing out the $\Psi_2$-component in the region $-R/2 \lesssim z \lesssim R/2$.
For $\Delta\Theta\neq 0$, the junction causes a superfluid current along the $z$-axis with a current velocity
 $v_\bot \sim \frac{\hbar}{m}\Delta\Theta/R>0$ or $v_\bot  \sim\frac{\hbar}{m}(\Delta\Theta-2\pi)/R<0$ [see Fig.~\ref{fig:Schematic}(b)].
If the current velocities through the two neighbouring junctions are parallel, the annihilation is completed between the junctions. On the other hand, the two junctions with opposite velocities leave a single-quantum vortex in the $\Psi_1$-component, whose core is filled with the $\Psi_2$-component.

The vortex formation strongly depends on how the junction grows from initial fluctuations. Although the growth rates of junctions with $v_\bot >0$ and $v_\bot <0$ are generally different, they are statistically equivalent for $\Delta\Theta = \pi$. Then, the junctions grow in a random fashion from initial random fluctuations, developing into the meshed structures shown in Fig.~\ref{fig:Schematic}(c). Vortices emerge as serpentine curves along the boundary between the two opposite junctions.

\subsection*{Method of numerical simulation}
\label{MethodNum}
Here, we briefly explain the methods of our numerical experiments. The sequence of the numerical computations of time development is as follows:
(i) A state with a brane--anti-brane pair is obtained by numerically minimizing the energy functional $\int d^3x({\cal K}+{\cal V})$ under the boundary condition $\Psi_1 =0$ at $z=0$ with $\arg\Psi_1=\pm \Delta\Theta/2$ for $z \lessgtr 0$.
(ii) The initial state of the time development is prepared by adding a random noise to the stationary solution.
(iii) The time development from the initial state is calculated by numerically integrating the equations of motion for the action $S$, called the coupled Gross-Pitaevskii equations,
\begin{eqnarray*}
i \hbar \partial_t \Psi_j =-\frac{\hbar^2}{2m}{\bm \nabla}^2\Psi_j+\sum_k g_{jk}|\Psi_k|^2\Psi_j-\mu_j\Psi_j.
\label{eq:GP}
\end{eqnarray*}
The numerical integrations are performed under the periodic boundary condition at $x,~y=\pm 51.2\xi$ and the Neumann boundary condition at $z=\pm 25.6 \xi$.
 The linear stability of a stationary solution $\Phi_j(z)$ of a brane--anti-brane pair is investigated by numerically diagonalizing the Bogoliubov-de Gennes equations, which are obtained by linearizing the Gross-Pitaevskii equations \eqref{eq:GP} with respect to a collective excitation $\delta \Psi_j ({\bm q})=\Psi_j({\bm r},t)-\Phi_j(z)=e^{\Gamma_q t}\left[u(z) e^{i{\bm q}\cdot{\bm r}-i\omega_q t}-v^*(z) e^{-i{\bm q}\cdot{\bm r}+i\omega_q t}\right]$ \cite{Pethick:2002}.

\subsection*{Coefficients of the tachyon potential}
 Here we show the coefficients $F_n$ of the tachyon potential $V=\frac{\mu_1^2\xi_\bot}{g}\sum_{n=0}^4 F_n{T}^n$ for the ansatz used in the text.
 The coefficients $F_n$ ($n=0,1,2,3,4$) are written as
 $F_n=A_n+B_n\theta(n_2)$ with
\begin{eqnarray*}
&&
A_4=\frac{16}{35},
\\
&&
A_3=\frac{2}{3} \cos \frac{\Delta \Theta }{2},
\\
&&
A_2=\frac{8}{45} \left(2 \gamma +3 \cos ^2\frac{\Delta \Theta }{2}-6 \nu  -2\right),
\\
&&
A_1=-\frac{1}{3} (2 \gamma -6 \nu +7) \cos \frac{\Delta \Theta }{2},
\\
&&
A_0=\frac{2}{9} (2 \gamma -6 \nu +7),
\\
&&
B_4=-\frac{32 \gamma ^2}{75},
\\
&&
B_3=-\frac{4}{5} \gamma ^2 \cos \frac{\Delta \Theta }{2},
\\
&&
B_2=
-\frac{3}{8} \gamma ^2 \cos ^2\frac{\Delta \Theta }{2}-\frac{32 \gamma
   ^2}{45}+\frac{32 \gamma  \nu }{15}-\frac{16 \gamma }{45},
\\
&&
B_1=
-\frac{1}{3} \gamma  (2 \gamma -6 \nu +1) \cos \frac{\Delta \Theta }{2},
\\
&&
B_0=
-\frac{2}{27} (2 \gamma -6 \nu +1)^2,
\end{eqnarray*}
where we used
\begin{eqnarray*}
&&
\xi_\bot=\left(\frac{2}{3}\gamma-2\nu+\frac{4}{3}\right)^{-1/2}\xi
\\
&&
n_2=\left[{\rm Re}\sqrt{
-\frac{4 \gamma}{5}T^2
-\frac{3\gamma}{4} T \cos \frac{\Delta \Theta }{2}
-\frac{2 \gamma }{3}+2 \nu-\frac{1}{3}
}\right]^2\frac{\mu_1}{g}.
\end{eqnarray*}
 The conditions  $\xi_\bot \geq \xi$ and $n_2^0\equiv n_2(T=0) \geq 0$ are reduced to
 $\frac{1}{3}\gamma+\frac{1}{6} \leq \nu \leq \frac{1}{3}\gamma+\frac{2}{3}$.
 The length parameter is fixed as $\xi_\bot=\xi$ for $\nu < \frac{1}{3}\gamma+\frac{1}{6}$ since the $\Psi_2$-component vanishes with $n_2^0=0$.

 The tachyon mass $m_T$ is written as
\begin{eqnarray*}
m_T^2 =-\frac{16}{45}(1+2\gamma^2+3\nu-6\gamma\nu)\frac{\mu_1^2\xi_\bot}{g}.
\end{eqnarray*}
 The instability condition $m_T^2<0$ is satisfied for $1+2\gamma^2+3\nu-6\gamma\nu>0$.
 The instability tends to be weak as $\nu$ increases, which is consistent with the numerical results of the linear stability analysis.
 This represents that with increasing $\nu$ the inter-brane distance $R\sim\xi_\bot$ increases, so that the inter-brane interaction becomes weak.
 Remember that our ansatz is not intended for the case of $R$ much larger than the brane thickness.
 In fact, the inter-brane distance $R$ approaches infinity for $\nu \to 1$ in the numerical results
 although one obtains $R =2 {\rm arcsinh}\left(\sqrt{gn_2/\mu_1}\right)\xi_\bot=\sqrt{6} {\rm arcsinh}\left(\sqrt{1/3}\right)\xi$ for $\nu = 1$ in our analytic model.

\subsection*{Surface tension of a kink}
We consider a flat kink parallel to the $y$-axis in equilibrium in the case of $\Delta \Theta=\pi$ with $V(-T)=V(T)$.
 The kink profile is the solution of the equation
\begin{eqnarray*}
0=\frac{\delta E_{2D}}{\delta T}=\frac{dV}{dT}+\frac{dG}{dT}(\partial_x T)^2\xi^2-2\partial_x\left[G(\partial_x T)\right]\xi^2.
\end{eqnarray*}
Integrating this equation once, and imposing the boundary conditions $T(x=0)=0$ and $T(\pm \infty)=\pm T_b$, one obtains
\begin{eqnarray*}
\xi\partial_x T=\sqrt{\frac{V(T)-V(T_b)}{G(T)}}.
\end{eqnarray*}
 This result can be used to evaluate the energy per unit length of kink, that is the surface tension, as
\begin{eqnarray*}
\sigma
&=&
\int^{+\infty}_{-\infty}dx\left[G(T)(\xi\partial_x T)^2+V(T)-V(T_b)\right]
\\
&=&
4\int^{+\infty}_{0}dx\left[V(T)-V(T_b)\right]
\\
&=&
4\xi\int^{T_b}_{0} dT \sqrt{G(T)\left[V(T)-V(T_b)\right]}.
\end{eqnarray*}
The coefficient $G$ is written as
\begin{eqnarray*}
G=\left[
\frac{2}{3}
+\frac{9}{64}\frac{\mu_1\gamma^2}{gn_2}\left(\frac{32}{15}T+\cos\frac{\Delta\Theta}{2}\right)^2
\theta(n_2)
\right]\frac{\mu_1^2\xi_\bot}{g},
\end{eqnarray*}
 where we used a function $\theta(n_2)=1$ for $n_2>0$ and $\theta(n_2)=0$ for $n_2=0$.
 Then the field $T(x)$ is non-smooth at $x=\pm x_0$, where $n_2 = 0$ for $|x|\geq x_0 \geq 0$.
 The field $T$ should become smooth if the gradient energy of $\Psi_2$-component is treated appropriately.
 However, a difference due to this unphysical non-smoothness would not affect on our order estimation.

 \subsection*{Defect nucleation rate}

\begin{figure} [hbtp] \centering
  \includegraphics[width=1. \linewidth]{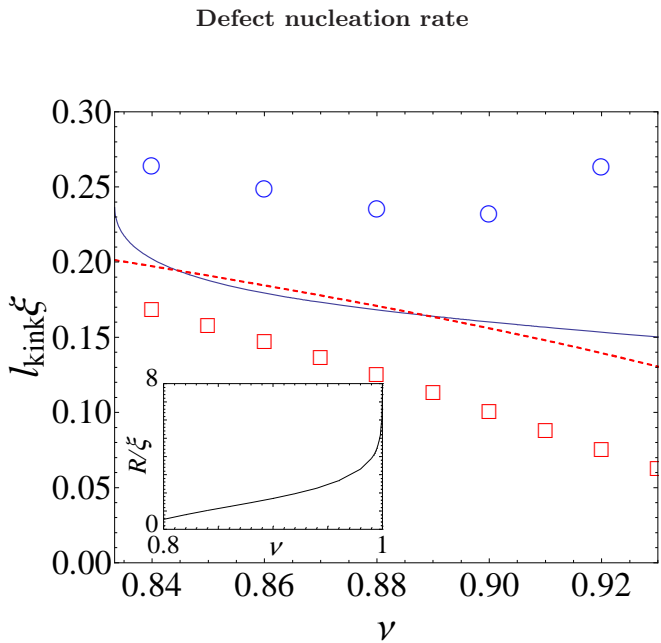}
  \caption{ Defect nucleation rates.
Vortex nucleation rates estimated through different analyses: the analytical estimation with Eq.\eqref{eq:Est_tension} (solid curve) and the numerical (squares) and analytical (dashed curve) estimations with Eq.~\eqref{Est_linear} based on the linear stability analysis. These estimations are compared with the projected vortex core density $l_{\rm 2D}$ (circles) at the same time as obtained from the numerical results shown in Fig.~\ref{fig:DecayKink}. The inset shows the numerical result of the $\nu$-dependence of the inter-brane distance $R$.
}
\label{fig:l_v}
\end{figure}
We can evaluate the vortex nucleation via the brane annihilation based on the effective tachyon field theory. In the following, we confine ourselves to $\Delta \Theta = \pi$. A vortex `projected' on the effective 2D system corresponds to a kink in the field $T$. When vortices are localized on the $z=0$ plane immediately after vortex nucleation, the total length $L_v$ of vortices in the original 3D system equals that of the kinks in the 2D system. The potential energy decreases when the tachyon field grows from $T=0$ toward the potential minima $T=\pm T_b$. The total energy can be conserved in the process of tachyon condensation if the decrease in the potential energy is compensated by the increase in the gradient energy by nucleating the kinks. Because the energy density in the initial state is $V(0)$ in the 2D system, the line density $l_{\rm kink}$ of the kink can be written as
\begin{eqnarray}
 l_{\rm kink}=\frac{L_v}{A}\sim\frac{V(0)-V(T_b)}{\sigma}
 \label{eq:Est_tension}
\end{eqnarray}
with the 2D system's area $A$ and the tension $\sigma$ (energy per unit length) of the kink.
Note that the conservation of $N_1$ and $N_2$ is not considered in the estimation explicitly.
Exactly speaking, in the process of vortex formation described by the variational ansatz,
 $N_1$ and $N_2$ are typically increased and decreased with time, respectively.
A simplest form of the ansatz is introduced so as to describe qualitatively the annihilation starting from a small inter-brane distance in strongly segregated condensates.
In that case, the changes of $N_1$ and $N_2$ are small in the process described with the ansatz,
 which causes a minor correction to the rate estimation.

 A kink solution of the field $T(x)$ parallel to the $y$-axis satisfies the equation $\delta E_{\rm 2D}/\delta T=0$ with the boundary condition $T\to \pm T_b$ for $x\to \pm \infty$, and the tension is calculated as
\begin{eqnarray}
\sigma=4\xi\int^{T_b}_{0} dT \sqrt{G(T)\left[V(T)-V(T_b)\right]}.
\label{eq:tension}
\end{eqnarray}
The line density $l_{\rm kink}$ is a decreasing function of $\nu$ because the tension $\sigma$ increases with $\nu$ more rapidly than $V(0)-V(T_b)$. The increase in the kink tension $\sigma$ in the 2D system may correspond to the increase in the vortex tension in the original 3D system, where the occupancy of the $\Psi_2$-component in the vortex cores of the $\Psi_1$-component is increased.

To demonstrate the validity of the effective model, the analytical estimation above is compared to the numerical results based on the linear stability analysis of the original 3D theory. We investigated the stability around the initial state against a fluctuation $\delta \Psi_j=e^{\Gamma_q t}\left[u(z) e^{i{\bm q}\cdot{\bm r}-i\omega_q t}-v^*(z) e^{-i{\bm q}\cdot{\bm r}+i\omega_q t}\right]$ in the form of the Bogoliubov excitation with growth rate $\Gamma_q$, wavenumber ${\bm q}$, and frequency $\omega_q$. The $q$-dependence of $\Gamma_q$ was investigated in a previous work \cite{Takeuchi:2011}.
For small $q$, $\Gamma_q$ increases linearly with $q$ and reaches its maximum $\Gamma_{\rm max}$ at $q=q_{\rm max}$.
A further increase in $q$ leads to decreasing $\Gamma_q$, which becomes zero at $q=q_{\rm top}$. The frequency $\omega_q$ is zero for $\Gamma_q\neq 0$.
 The values $\Gamma_{\rm max}$, $q_{\rm max}$, and $q_{\rm top}$ decrease as $\nu$ increases. If the instability is sufficiently strong, {\it i.e.} $\Gamma_{\rm max}$ is large, the line density $l_{\rm kink}$ can be estimated as
\begin{eqnarray}
 l_{\rm kink} \sim\frac{q_{\rm max}}{\pi}.
\label{Est_linear}
\end{eqnarray}
The numerical estimation with Eq.~\eqref{Est_linear} (squares) is reasonably consistent with the analytical estimation with Eq.~\eqref{eq:Est_tension} (solid curve) in Fig.~\ref{fig:l_v}. Both estimations are invalid for $\nu\to 1$. This is because the numerical results show that $R$ diverges and $\Gamma_{\rm max}$ approaches zero for $\nu\to 1$ [see the inset of Fig.~\ref{fig:l_v}], where both the effective theory and the linear stability analysis are inapplicable.

It is instructive to construct an equation of motion for the rolling tachyon from the effective field theory. The equation of motion conventionally contains a term proportional to
$\delta E_{2D}/\delta T\propto (-G(0)\xi^2{\bm \nabla}_\parallel^2+m_T^2)T $ around $T=0$.
 By considering a small perturbation $\propto e^{\Gamma_q t}\sin (qx)$ with $\Gamma_q\propto q$ for $q\to 0$, a dimensional analysis gives
\begin{eqnarray}
\hbar^2\partial_t^2 T \sim \frac{g}{\xi_\bot}\xi^2{\bm \nabla}_\parallel^2(-G(0)\xi^2{\bm \nabla}_\parallel^2+m_T^2)T~~~~(T \ll 1).
\label{eq:Tmotion}
\end{eqnarray}
Then, one obtains
\begin{eqnarray}
\hbar \Gamma_q=\xi q \sqrt{\frac{g}{\xi_\bot}\left[G(0)\xi^2q^2-m_T^2\right]},
\end{eqnarray}
and
\begin{eqnarray}
q_{\rm max}\xi \sim \sqrt{-\frac{m_T^2}{2G(0)}},
\end{eqnarray}
which gives a quantitative agreement with the defect nucleation rate, as shown in Fig.~\ref{fig:l_v}(b) (dashed curve). For $q\to 0$, the equation \eqref{eq:Tmotion} yields a Klein--Gordon-type equation
$\partial_t^2 T\sim c_T^2{\bm \nabla}_\parallel^2T$, which describes a massless particle with an imaginary phase velocity $c_T=\sqrt{gm_T^2/m\mu_1\xi_\bot}$.
This particle has a growth rate for zero momenta, $\Gamma_{q}=0$, as a consequence of the energy conservation in the growth process. We may construct an equation of motion similarly around the potential minima $T=\pm T_b$, and then, the equation describes a massless particle with a real phase velocity. However, the minima $T=\pm T_b$ should be distinct from the true vacuum of this system, because the phase gradient $\partial_z \arg\Psi_1$ is localized around $z=0$. In the true vacuum, $\Psi_1={\rm const.}$ and $\Psi_2=0$, obtained after the complete annihilation, the dispersion of perturbations has in turn a real phase velocity $\sqrt{\mu_1/m}$ with $|\Psi_1|=\sqrt{\mu_1/g}$ for the massless particle, that is, a phonon.


\subsection*{Decay of projected vortex line density}
\begin{figure} [htbp] \centering
  \includegraphics[width=1. \linewidth]{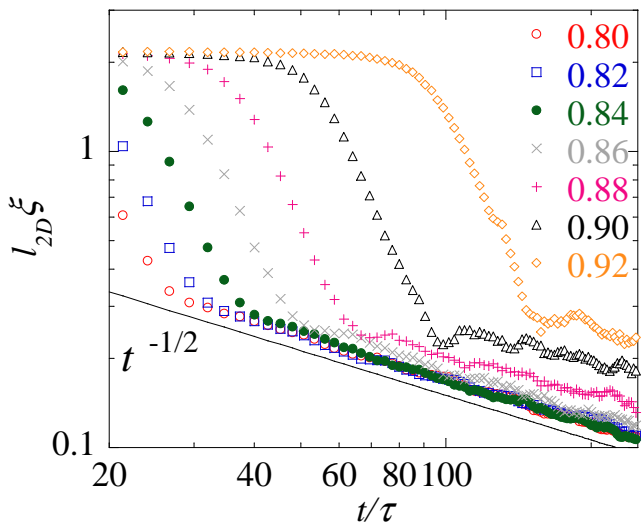}
  \caption{ Decay of the projected vortex core density.
 Different marks show the plots of the density $l_{\rm 2D}$ for $\nu=0.80$, $0.82$, $0.84$, $0.86$, $0.88$, $0.90$, and $0.92$. Each plot obeys a power law $t^{-1/2}$ when the structure factor is close to the scaling function, implying that the projected field behaves as a non-conserved field. The values of $l_{\rm 2D}$ at time $t=t_v$, after which $l_{\rm 2D}$ obeys the power law, are compared to the estimations of the defect nucleation rate in Fig.~\ref{fig:l_v}. For large $\nu$, the value $l_{\rm 2D}(t_v)$ increases unphysically and the plots fluctuate strongly because many ghost vortices remain because of a large amount of the $\Psi_2$-component.
}
\label{fig:DecayKink}
\end{figure}

The time dependence of the projected defect density $l_{\rm 2D}$ also contains information that is important for understanding the system.
 The phase ordering kinetics suggests the growth law $1/l_{\rm 2D}\propto t^{1/3}$ or $1/l_{\rm 2D}\propto t^{1/2}$ depending on whether the order parameter field is conserved or not, respectively \cite{Bray:1994}. The decay of the density $l_{\rm 2D}$ follows a power law, $1/l_{\rm 2D}\propto t^{1/2}$, in a manner similar to the different values of $\nu$ as shown in Fig.~\ref{fig:DecayKink}. This indicates that the effective energy in the projected system dissipates in the projected phase ordering dynamics. To explain this dissipation thoroughly, we have to consider some degrees of freedom that were neglected in our analysis, such as the motions of the $\Psi_2$-component along the vortex core, emissions of vortex rings, and density pulses to the extra dimension induced by vortex reconnection \cite{Ex.1}.



\begin{thebibliography}{99}

\bibitem{Sen:2005}
A. Sen,
 Int. J. Mod. Phys. A {\bf 20}, 5513-5656 (2005).

\bibitem{Polchinski:1998}
J. Polchinski,
{\it String Theory} (Cambridge University Press, Cambridge, England, 1998), vols. 1 and 2, chap. 13.

\bibitem{Dvali:2001}
G. Dvali, Q. Shafi, and S. Solganik,
arXiv:hep-th/0105203.

\bibitem{Langlois:2002}
D. Langlois,
 Prog. Theor. Phys. Suppl. {\bf 148}, 181-212 (2002).

\bibitem{Quevedo:2002}
F. Quevedo,
 Class. Quantum Grav. {\bf 19}, 5721 (2002)

\bibitem{McAllister:2008}
L. McAllister and E. Silverstein,
 Gen. Rel. Grav. {\bf 40}, 565-605 (2008).


\bibitem{Jones:2002}
N. Jones, H. Stoica, and S.-H. H. Tye,
 J. High Energy Phys. {\bf 07}, 051 (2002).

\bibitem{Sarangi:2002}
S. Sarangi, and S.-H. H. Tye,
Phys. Lett. B {\bf 536}, 185-192 (2002).

\bibitem{Dvali:2004}
G. Dvali, and A. Vilenkin,
J. Cosmol. Astropart. Phys. {\bf 03} (2004) 010. 



\bibitem{Kibble:1976}
T. W. B. Kibble, 
J. Phys. A {\bf 9}, 1387-1398 (1976).

\bibitem{Zurek:1985}
W. H. Zurek,
Nature {\bf 317}, 505-508 (1985);
Phys. Rep. {\bf 276}, 177-221 (1996).

\bibitem{Vilenkin:1994}
A. Vilenkin and E. P. S. Shellard,
{\it Cosmic Strings and Other Topological Defects}
 (Cambridge University Press, Cambridge, England, 1994).




\bibitem{Hendry:1994}
P. C. Hendry,
N. S. Lawson, R. A. M. Lee, P. V. E. Mcclintock, and C. D. H. Williams,
Nature {\bf 368}, 315-317 (1994).

\bibitem{Bowick:1994}
Mark J. Bowick, L. Chandar, E. A. Schiff and Ajit M. Srivastava,
Science {\bf 263}, 943-945 (1994).

\bibitem{Bauerle:1996}
C. B\"{a}uerle, Yu. M. Bunkov, S. N. Fisher, H. Godfrin, G. R. Pickett,
Nature {\bf 382}, 332-334 (1996).

\bibitem{Ruutu:1996}
V. M. H. Ruutu, V. B. Eltsov, A. J. Gill, T. W. B. Kibble, M. Krusius, Yu. G. Makhlin, B. Placais, G. E. Volovik,and Wen Xu,
Nature {\bf 382}, 334-336 (1996).

\bibitem{Carmi:2000}
R. Carmi, E. Polturak, and G. Koren,
Phys. Rev. Lett. {\bf 84}, 4966-4969 (2000).

\bibitem{Maniv:2003}
A. Maniv, E. Polturak, and G. Koren,
Phys. Rev. Lett. {\bf 91}, 197001 (2003).

\bibitem{Monaco:2006}
R. Monaco, J. Mygind, M. Aaroe, R. J. Rivers, and V. P. Koshelets,
Phys. Rev. Lett. {\bf 96}, 180604 (2006).


\bibitem{Sadler:2006}
L. E. Sadler, J. M. Higbie, S. R. Leslie, M. Vengalattore, and D. M. Stamper-Kurn,
 Nature {\bf 443}, 312-315 (2006).

\bibitem{Weiler:2008}
Chad N. Weiler, Tyler W. Neely, David R. Scherer, Ashton S. Bradley, Matthew J. Davis, and Brian P. Anderson,
 Nature {\bf 455}, 948-951 (2008).


\bibitem{Anderson:2001}
B. P. Anderson, P. C. Haljan, C. A. Regal, D. L. Feder, L. A. Collins, C.W. Clark, and E. A. Cornell,
Phys. Rev. Lett. {\bf 86}, 2926-2929 (2001).


\bibitem{Kasamatsu:2010}
 K. Kasamatsu, H. Takeuchi, M. Nitta, and M. Tsubota,
 J. High Energy Phys.  {\bf 11}, 068 (2010).
 A similar brane configuration for a spin-1 BEC has been reported in M. O. Borgh and J. Ruostekoski, Phys. Rev. Lett. {\bf 109}, 015302 (2012).


\bibitem{Takeuchi:2011}
H. Takeuchi, K. Kasamatsu, M. Nitta, and M. Tsubota,
J. Low Temp. Phys. {\bf 162}, 243-249 (2011).

\bibitem{Bray:1994}
A. J. Bray,
 Adv. Phys. {\bf 43}, 357-459 (1994).

\bibitem{Bradley:2008}
D. I. Bradley, S. N. Fisher, A. M. Guenault, R. P. Haley, J. Kopu, H. Martin, G. R. Pickett, J. E. Roberts and V. Tsepelin,
 Nat. Phys. {\bf 4}, 46-49 (2008).




\bibitem{Pethick:2002}
C. J. Pethick, and H. Smith,
 {\it Bose-Einstein Condensation in Dilute Gases}
 (Cambridge University Press, Cambridge, England, 2008), 2nd ed.


\bibitem{Papp:2008}
S. B. Papp, J. M. Pino, and C. E. Wieman,
 Phys. Rev. Lett. {\bf 101}, 040402 (2008).


\bibitem{Supp} See SUPPLEMENTAL MATERIAL for supplemental data and discussions.

\bibitem{Foot3} This assumption is intended to simulate the phase ordering dynamics after a rapid quench in the two-dimensional subspace described later.

\bibitem{Foot4} To focus our discussion mainly on similar situations in the experiment \cite{Anderson:2001},
 we used this ansatz with a discontinuous derivative $\partial_z\Psi_1(t=0)$ at $z=0$,
 which is realized as a solution under the boundary condition $\Psi_1(z=0)=0$.
\bibitem{Foot1} The squared `mass' becomes positive in the stable vacuum obtained after the instability.
In the effective theory for $\Delta\Theta=\pi$, the stable vacuum corresponds to a minimum, $T=T_b$ or $T=-T_b$, as seen in Fig. \ref{fig:Tpotential} (a).

\bibitem{Landau:1980StatisticalPhysics} See, for example, L. D. Landau and E. M. Lifshitz, {\it Statistical Physics: Course of Theoretical Physics} (Pergamon, New York, 1980), 3rd ed., vol. 5, part 1, chap. 14. 

\bibitem{Foot2} The effective energy $E_{\rm 2D}$ in the projected-2D system is invariance under inversion $T=-T$ for $\Delta\Theta =\pi$.
 The unstable vacuum (the initial state at $t=0$) with zero `magnetization' ($T=0$) is invariant under the transformation $T=-T$.
 This $Z_2$ symmetry is spontaneously broken in the stable vacuum ($T=T_b$ or $T=-T_b$) of the effective theory.

\bibitem{Nitta:2012}
M. Nitta, K. Kasamatsu, M. Tsubota, and H. Takeuchi,
Phys. Rev. A {\bf 85}, 053639 (2012).









\end{thebibliography}

\begin{thebibliography}{99}

\bibitem[S1]{Pethick:2002}
C. J. Pethick, and H. Smith,
 {\it Bose-Einstein Condensation in Dilute Gases, 2nd ed.}
 (Cambridge University Press, Cambridge, 2008).


\bibitem[S2]{Takeuchi:2011}
H. Takeuchi, K. Kasamatsu, M. Nitta, and M. Tsubota,
J. Low Temp. Phys. {\bf 162}, 243-249 (2011).



\bibitem[S3]{Bray:1994}
A. J. Bray,
 Adv. Phys. {\bf 43}, 357-459 (1994).
 
\bibitem[S4]{Ex.1}
Such motions become more pronounced after vortices are clearly formed,
 but are less effective before vortices emerge and just after the vortices form as can be seen from the numerical simulation of Fig.~1 (c) in the text (compare the plots $t/\tau=40$ and $t/\tau=56$).
This is why the defect nucleation rate of Eq.~\eqref{eq:Est_tension} is estimated by assuming that the total energy is conserved.

\end{thebibliography}
\end{document}